\documentclass[aps,prl,twocolumn,showpacs]{revtex4-1}
\usepackage{dcolumn}
\usepackage{bm}
\usepackage{graphicx}
\usepackage{amsmath}
\usepackage{latexsym}
\usepackage{amsfonts}
\usepackage{amssymb}
\usepackage{array}
\usepackage{epsfig}
\usepackage{times}
\usepackage{subfigure}
\usepackage{bbm}

\newcommand{\projsm}[2]{\vert#1\rangle\langle#2\vert}

\newcommand{\pare}[1]{\left(#1\right)}
\newcommand{\beqn}{\begin{eqnarray}}
\newcommand{\eeqn}{\end{eqnarray}}
\newcommand{\ket}[1]{\left\vert#1\right\rangle}

\newcommand{\proj}[2]{\left\vert#1\rangle\langle#2\right\vert}

\newcommand{\be}{\begin{equation}}
\newcommand{\ee}{\end{equation}}
\newcommand{\ba}{\begin{array}}
\newcommand{\ea}{\end{array}}
\newcommand{\bea}{\begin{eqnarray}}
\newcommand{\eea}{\end{eqnarray}}

\newcommand{\bqa}{\begin{eqnarray}}
\newcommand{\eqa}{\end{eqnarray}}
\DeclareSymbolFont{symbols}{OMS}{cmsy}{m}{n}

\begin{document}

\title{Quantum Simulation of the Ultrastrong Coupling Dynamics in Circuit QED}
\author{D. Ballester,$^1$ G. Romero,$^1$ J. J. Garc\'ia-Ripoll,$^2$ F. Deppe,$^{3,4}$ and E. Solano$^{1,5}$ }
\affiliation{$^1$Departamento de Qu\'imica F\'isica, Universidad del Pa\'is Vasco UPV/EHU, Apartado 644, 48080 Bilbao, Spain\\
$^2$Instituto de F\'isica Fundamental, CSIC, Serrano 113-bis, 28006 Madrid, Spain\\
$^3$Walther-Mei\ss ner-Institut, Bayerische Akademie der Wissenschaften, D-85748 Garching, Germany\\
$^4$Physik-Department, Technische Universit\"at M\"unchen, D-85748 Garching, Germany\\
$^5$IKERBASQUE, Basque Foundation for Science, Alameda Urquijo 36, 48011 Bilbao, Spain}

\date{\today}
   
\begin{abstract} 
We propose a method to get experimental access to the physics of the ultrastrong (USC) and deep strong (DSC) coupling regimes of light-matter interaction through the quantum simulation of their dynamics in standard circuit QED. The method makes use of a two-tone driving scheme, using state-of-the-art circuit-QED technology, and can be easily extended to general cavity-QED setups. We provide examples of USC/DSC quantum effects that would be otherwise inaccessible.
\end{abstract}

\maketitle

\paragraph{Introduction.}

The Jaynes-Cummings model (JCM) \cite{JC} is a cornerstone of the field of quantum optics. It describes the interaction between a quantized electromagnetic (EM) field mode and a two-level atom under two important assumptions. First, the interaction is accurately modeled by a dipolar coupling between the field and the atom. Second, one can apply the rotating-wave approximation (RWA) because the coupling is small enough when compared to the sum of frequencies of the two-level atom and EM field. These restrictions yield a solvable model where atom and field exchange one excitation.

To study experimentally the physics of the JCM, the interaction needs to reach the strong-coupling (SC) regime. This can be done by isolating the two-level system from free space by means of a cavity with highly reflecting mirrors, making the coupling strength much larger than the spontaneous emission rate and the cavity decay rate. This type of setup is known as cavity quantum electrodynamics (cavity QED) \cite{Mabuchi, Haroche}. Many relevant features of the JCM have been successfully tested in actual experiments using cavity-QED technology. For instance, the observation of the vacuum Rabi mode splitting in the optical domain with Alkali atoms was reported \cite{Kimble}, while in the microwave regime, vacuum Rabi oscillations using Rydberg atoms have also been realized \cite{Haroche, Brune}.

In 2004, an important step forward was made when an artificial two-level atom (or a qubit) was shown to be strongly coupled to the EM field inside a superconducting 1D transmission-line resonator~\cite{Blais}. The newly born circuit QED technology was rapidly recognized as a promising architectural platform from which the study of light-matter interaction can be extended \cite{ReviewCircuit}. Although most circuit QED implementations were restricted to the SC regime of the JCM, key experiments showing the breakdown of the RWA have been recently realized in semiconductor microcavities \cite{Ciuti} and circuit QED~\cite{Niemczyk, FD}. They have opened up new directions of research into the ultrastrong coupling (USC) regime of light-matter interaction~\cite{Ciuti-th,Bourassa}, where the RWA can no longer be used, leading to novel features such as the creation of photons from the quantum vacuum \cite{Ciuti-th}. Even though these works show that reaching the ultrafast dynamics is feasible, its controllability becomes very demanding as the light-matter coupling increases~\cite{Peropadre}.

In this work, we introduce the quantum simulation of the USC/DSC dynamics in circuit QED with a qubit-cavity system in the SC regime. Our treatment makes use of a novel two-tone orthogonal driving to the qubit. We show through analytical and numerical calculations that our proposal will have access to the regimes of USC ($0.1\lesssim \! g/ \omega \lesssim \! 1$, $g/ \omega$ being the ratio of the coupling over the resonator frequency) and DSC~\cite{DSC} ($g/ \omega\gtrsim \! 1$). This will pave the way for the implementation of a quantum simulator \cite{ReviewQS} for a wide range of regimes of light-matter coupling \cite{Braak} in systems where they are unattainable. As we will discuss below, this includes the simulation of relativistic quantum phenomena, Dicke/spin-boson model, Kondo physics, and Jahn-Teller instability~\cite{Duty}. Although we present our method in the language of circuit QED, it can also be realized in microwave cavity QED~\cite{Haroche,Sol}.

\begin{figure*}
\includegraphics[width=0.48\textwidth]{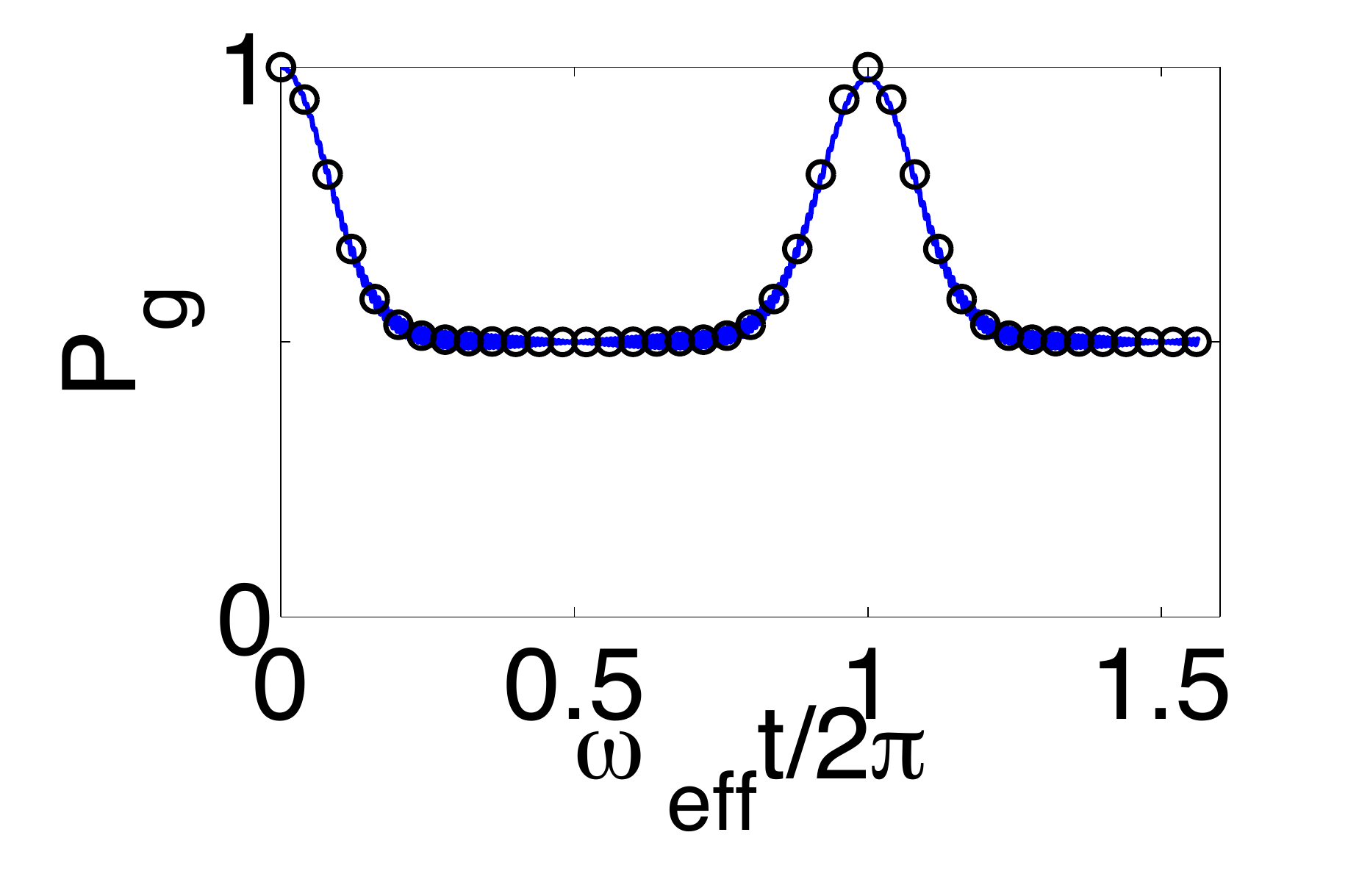}
\includegraphics[width=0.474\textwidth]{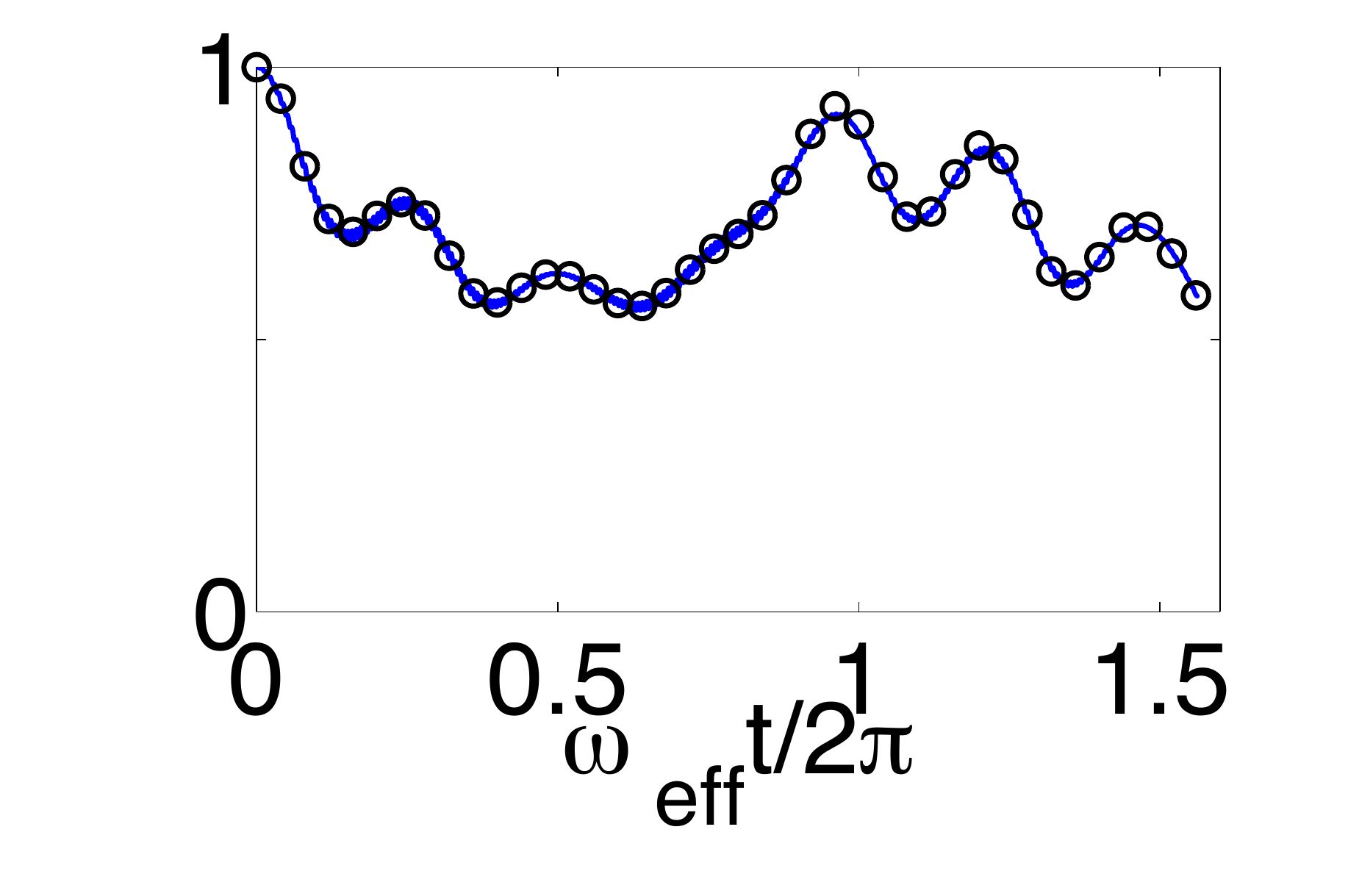}
\caption{\label{cqed:figs}
$P_{g}(t)$ obtained by integrating the exact (solid line) Eq.~(\ref{HamilDriv}) and the effective (circles) Hamiltonian Eq.~(\ref{HamilEff}). With the set of parameters in the text, $\Omega_1= 2\pi\times700\,$MHz, we have two cases: {\bf (left panel)} $\Omega_2  =0$;  {\bf (right panel)} $\Omega_2  =2\pi\times10\,$MHz. The simulated ratio is $g_{\rm eff} / \omega_{\rm eff} =1$.} 
\end{figure*}

\paragraph{The model.}

The physical system we consider consists of a superconducting qubit strongly coupled to a microwave resonator mode. If we work in the qubit degeneracy point, the Hamiltonian is given by \cite{Blais}
\begin{eqnarray}
{\cal H} =  \frac{\hbar \omega_q}{2} \sigma_z + \hbar  \omega a^\dag a  -\hbar  g    \sigma_x  \pare{a + a^\dag} \label{HamilDiag}, 
\end{eqnarray}
where $\omega_q$, $\omega$ are the qubit and photon frequencies, $g$ stands for the coupling strength: Likewise $a$($a^\dag$) represent the annihilation(creation) operators of the photon field mode, whereas $\sigma_x = \sigma^+ + \sigma = \projsm{e}{g}+ \projsm{g}{e}$, $\sigma_z = \projsm{e}{e}-\projsm{g}{g}$, being ${\ket{g},\ket{e}}$ the ground and excited eigenstates of the qubit. In a typical circuit-QED implementation, this Hamiltonian can be simplified further by applying the RWA. Precisely~\cite{Zueco}, if $\{|\omega-\omega_q|, g\} \ll\omega+\omega_q$, then it can be effectively approximated as
\begin{eqnarray}
{\cal H} &=& \frac{\hbar \omega_q}{2} \sigma_z +\hbar  \omega a^\dag a -\hbar  g \pare{\sigma^\dag a + \sigma a^\dag} ,\label{HamilRWA}
\end{eqnarray}
which resembles the celebrated JCM of quantum optics. Performing the RWA amounts to neglect counter-rotating terms $\sigma a$ and $\sigma^+ a^+$, leading to a Hamiltonian Eq.~(\ref{HamilRWA}) where the number of excitations is conserved.

This Hamiltonian (\ref{HamilRWA}) will be the starting point of our derivations. Consider that the qubit is now orthogonally driven by two classical fields. The Hamiltonian of the driven system reads
\begin{eqnarray}
  {\cal H} &=&  \frac{\hbar \omega_q}{2} \sigma_z +\hbar  \omega a^\dag a -\hbar  g \pare{\sigma^\dag a + \sigma a^\dag}  \nonumber \\  & & \hspace{-13mm} - \hbar \Omega_1  \pare{ e^{i \omega_1 t} \sigma + e^{-i \omega_1 t} \sigma^\dag} - \hbar  \Omega_2 \pare{ e^{i \omega_2 t} \sigma + e^{-i \omega_2 t} \sigma^\dag}, \label{HamilDriv}
\end{eqnarray}
where $\Omega_j$, $\omega_j$ stand for the amplitude and frequency of the $j-$th driving. Note that the orthogonal drivings couple to the qubit in a similar fashion as the resonator field does. To obtain (\ref{HamilDriv}), we have implicitly assumed that the RWA applies not only to the qubit-resonator interaction term, but also to both drivings. %
Next, we will write (\ref{HamilDriv}) in the reference frame rotating with the frequency of the first driving, $\omega_1$, that is,
\begin{eqnarray}
{\cal H}^{L_1} &=&\hbar  \frac{\omega_q-\omega_1}{2} \sigma_z +\hbar  (\omega-\omega_1) a^\dag a - \hbar g  \pare{\sigma^\dag a +\sigma a^\dag} \nonumber \\ & &  \hspace{-13mm} - \hbar \Omega_1  \pare{ \sigma + \sigma^\dag } - \hbar  \Omega_2   \pare{ e^{i (\omega_2-\omega_1) t} \sigma + e^{-i (\omega_2-\omega_1) t} \sigma^\dag } .
\end{eqnarray}
This allows us to transform the original first driving term into a time independent one ${\cal H}_0^{L_1} = - \hbar \Omega_1  \pare{ \sigma + \sigma^\dag} $, leaving the excitation number unchanged. We will assume this to be the most significant term and treat the others perturbatively by going into the interaction picture with respect to ${\cal H}_0^{L_1} $, ${\cal H}^{I} (t) = e^{i {\cal H}_{0}^{L_1} t/\hbar } \pare{{\cal H}^{L_1}   - {\cal H}_0^{L_1} }    e^{-i {\cal H}_{0}^{L_1} t/\hbar } $. The physical implementation of this transformation based on a Ramsey-like pulse is described later in the text. Using the rotated spin basis $\ket{\pm} = \pare{\ket{g} \pm \ket{e} }/\sqrt2$, we have
\begin{eqnarray}
{\cal H}^{I} (t)  \!  &=&  \!  -\hbar  \frac{\omega_q-\omega_1}{2}  \!\!  \pare{ e^{-i 2  \Omega_1  t}  \! \proj{+}{-}   \! +  \!   {\rm H.c.}} \!  + \hbar (\omega-\omega_1) a^\dag a \nonumber  \\
  & -& \frac{\hbar g }{2} \left( \left\{  \proj{+}{+} - \proj{-}{-} + e^{-i 2  \Omega_1 t} \proj{+}{-} \right. \right. \nonumber \\
 & & -  \left.\left. e^{i 2  \Omega_1  t}\proj{-}{+}  \right\} a + {\rm H.c.} \right) \nonumber \\  
  &-&  \frac{\hbar \Omega_2 }{2} \left(   \left\{  \proj{+}{+} - \proj{-}{-} - e^{-i 2  \Omega_1  t} \proj{+}{-}  \right. \right. \nonumber \\ & & \left. \left. + e^{i 2 \Omega_1  t}\proj{-}{+}  \right\} e^{i (\omega_2-\omega_1) t} + {\rm H.c.} \right). \label{HI1}
\end{eqnarray}
By tuning the parameters of the external drivings as $\omega_1-\omega_2=2 \Omega_1 ,$ we can choose the resonant terms in this time-dependent Hamiltonian. Then, in case of having a relatively strong first driving, $ \Omega_1$, we can approximate the expression above by a time-independent effective Hamiltonian as
\begin{eqnarray}
{\cal H}_{\rm eff}
=  \hbar (\omega-\omega_1) a^\dag a + \frac{\hbar\Omega_2}{2} \sigma_z   -  \frac{\hbar g}{2} \sigma_x \pare{a+a^\dag} .  \label{HamilEff}
\end{eqnarray}
Note the resemblance between the original Hamiltonian (\ref{HamilDiag}) and (\ref{HamilEff}). While the value of the coupling $g$ is fixed in (\ref{HamilEff}), we can still tailor the effective parameters by tuning amplitudes and frequencies of the drivings. If we can reach values such that $\Omega_2  \sim (\omega-\omega_1) \sim g/2$, the dynamics of the original system will simulate that of a qubit coupled to the resonator with a relative interaction strength beyond the SC regime---ideally in the USC/DSC. The coupling regime reproduced through the effective Hamiltonian (\ref{HamilEff}) can be quantified by the ratio $g_{\rm eff} / \omega_{\rm eff}$, where $g_{\rm eff} \equiv g/2$ and $ \omega_{\rm eff}\equiv\omega-\omega_1$.

\begin{figure*}
\includegraphics[width=0.48\textwidth]{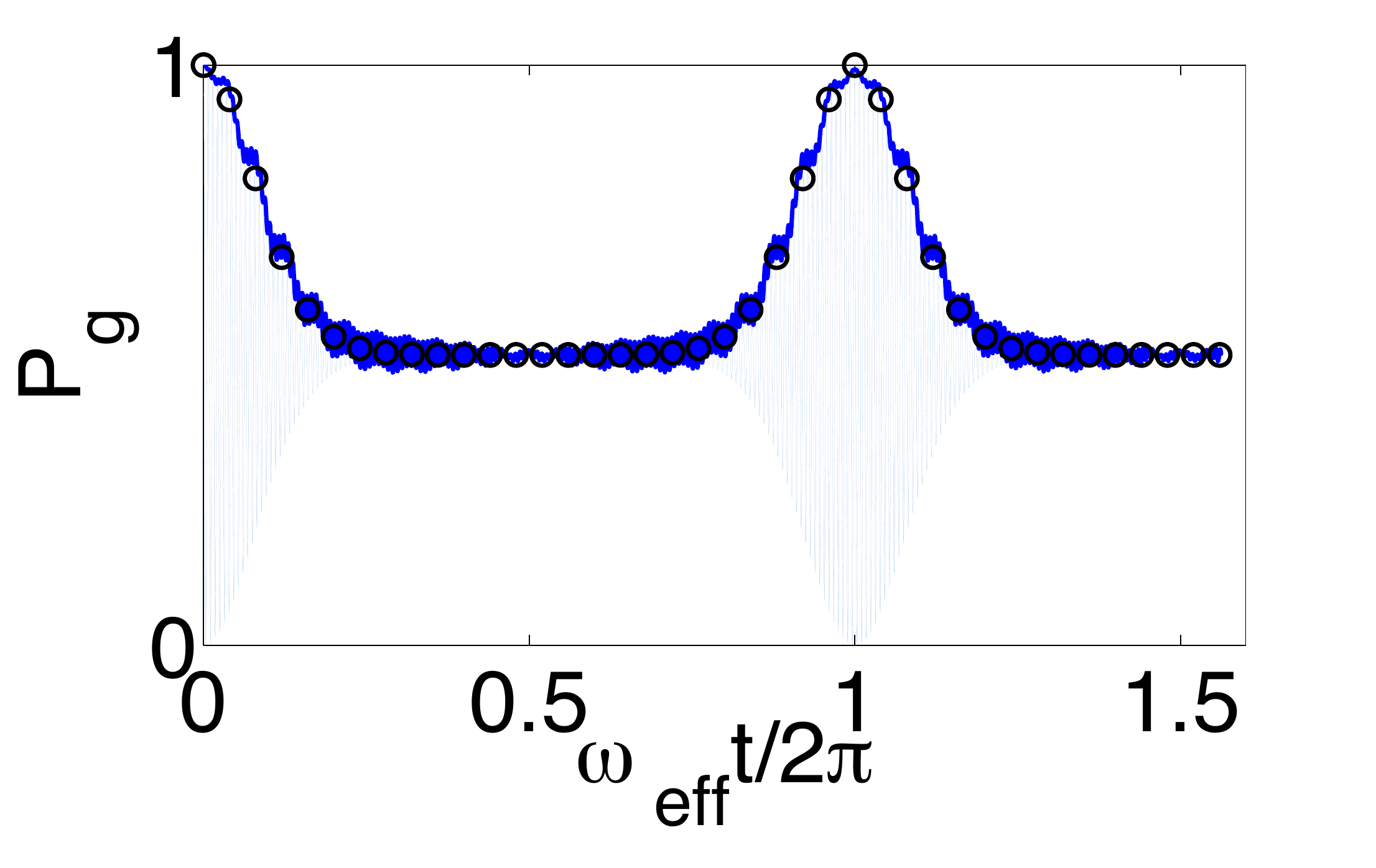}
\includegraphics[width=0.474\textwidth]{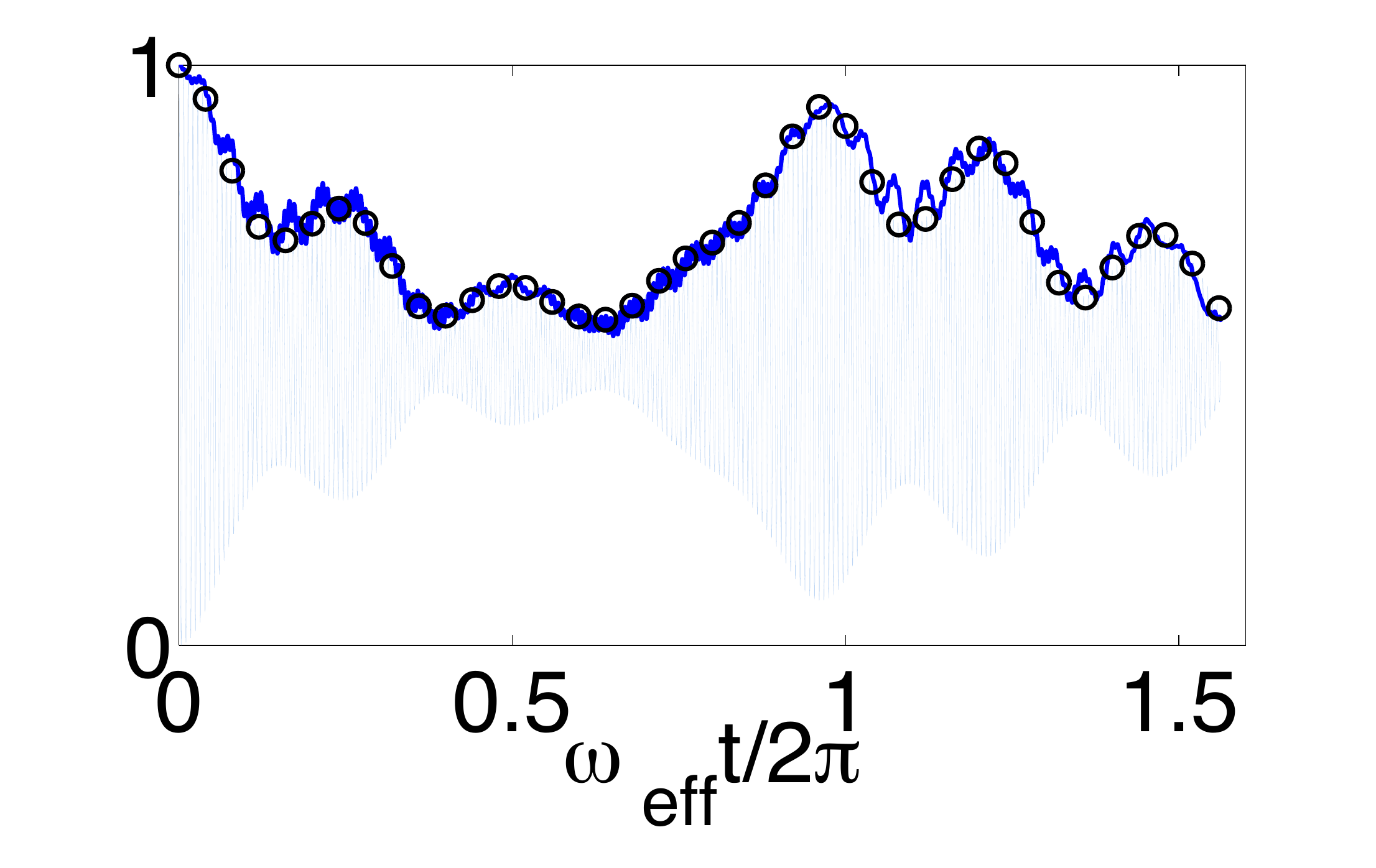}
\caption{\label{intpict:figs}
$P_{g}(t)$ obtained by integrating the exact Hamiltonian Eq.~(\ref{HamilDriv}) (light solid line, shaded area), and after applying a Ramsey-like pulse (dark solid). Both are compared to the effective (circles) Hamiltonian Eq.~(\ref{HamilEff}) evolution. The same parameters of Fig.~\ref{cqed:figs} is used: {\bf (left panel)} $\Omega_2  =0$;  {\bf (right panel)} $\Omega_2  =2\pi\times10\,$MHz. During the Ramsey-like pulse, qubit energy and driving frequency are detuned by $-2\pi\times200\,$MHz.} 
\end{figure*}

\paragraph{Numerical analysis.}

To study the feasibility of our proposal, we have performed numerical calculations with realistic parameters for circuit QED \cite{Blais} and compare the evolution described by the exact and effective Hamiltonians. In principle, the choice of $\Omega_1$ is unimportant as long as it is strong compared to the other parameters involved in this problem. Indeed, the evolutions of the Hamiltonians of Eqs.~(\ref{HamilDriv}) and (\ref{HamilEff}) become more similar with increasing $\Omega_1  $. Nonetheless, experimental restrictions are expected to set the limit of this driving strength. Throughout the rest of this work, we assume $\Omega_1\sim 2\pi\times0.7\,$GHz, which is
a realistic value.

After the discussion regarding the DSC dynamics \cite{DSC}, an interesting experiment would be the following. Assume we prepare the original SC undriven system in its ground state, i.e. $\ket{g,0}$, and then at time $t=0$ we switch on the external drivings. Now the system evolves according to the unitary operator which is computed by integrating the driven Hamiltonian Eq.~(\ref{HamilDriv}). For the sake of simplicity, we now assume that the evolution of the state is calculated in the rotating reference frame of Hamiltonian ${\cal H}^{I} (t)$. Later we will discuss how this step can be implemented. The solid line in Fig.~\ref{cqed:figs} shows the evolution of $P_{g}(t)$ for two different cases. Both show a very good agreement when compared to the same probability but computed from the effective Hamiltonian (circles) of Eq.~(\ref{HamilEff}) derived in the strong driving limit.  All these simulations are done with values of $\omega_q= 2\pi\times 8.01\,$GHz, $\omega= 2\pi\times8.01\,$GHz, $g =2\pi\times20\,$MHz, $\omega_1= 2\pi\times8\,$GHz, $\omega_2 =2\pi\times6.6\,$GHz, and $\Omega_1 =2\pi\times0.7\,$GHz. Clearly such an amplitude for the first strong driving suffices, even to simulate the dynamics of a system reaching $g_{\rm eff} / \omega_{\rm eff} =1$. A feature characteristic of the DSC dynamics is the existence of a well defined periodic evolution for the probability $P_{g}(t)$, in the case of a degenerate qubit. Starting from 1, $P_{g}(t)$ decays to 0.5 quite rapidly, to have a subsequent revival at a time that is equal to the inverse of the effective resonator frequency. This is accompanied by the generation of photon number wavepackets that oscillate in time \cite{DSC}. Putting $\Omega_2 =0$ (Fig.~\ref{cqed:figs}, left panel) in our simulation we observe that $P_{g}(t)$ presents nearly perfect revivals that take place at $\omega_{\rm eff} t_{\rm rev}\equiv g_{\rm eff} t_{\rm rev}=2\pi$, which corresponds to $t_{\rm rev}=0.1$ $\mu$sec for the set of parameters used. If the effective energy of the simulated qubit $\Omega_2 $ is not zero (Fig.~\ref{cqed:figs}, right panel) the evolution becomes nonperiodic, producing a distortion of the revival peack, which no longer reaches unity.

\paragraph{Ramsey-like pulse.}

The computation of the probability $P_{g}(t)$ for the exact Hamiltonian Eq.~(\ref{HamilDriv}) in Fig.~\ref{cqed:figs} has been done in the rotating reference frame used to derive Hamiltonian ${\cal H}^{I} (t)$. However, without going into this interaction picture, the evolution of $P_{g}(t)$ would show a fast oscillating term, depicted by the light solid line shown in Fig.~\ref{intpict:figs}. Here, we propose the following scheme based on a Ramsey-like pulse, in order to get rid of this strong oscillation in an experiment. Imagine that after letting the system evolve with the Hamiltonian of Eq.~(\ref{HamilDriv}) for a time $t$, we switch off non-adiabatically both external drivings and apply a detuning of about $-2\pi\times200\,$MHz to the frequency of the qubit, $\omega_q$, from its original value $2\pi\times 8.01\,$GHz. Next, we switch on a third driving with frequency detuned by $-2\pi\times200\,$MHz, from the value of $\omega_1= 2\pi\times8\,$GHz, and with amplitude $-\Omega_1$ (opposite phase relative to the first one). The application of this Ramsey-like pulse will take the same time $t$. The wiggly dark solid line of Fig.~\ref{intpict:figs} corresponds to the measurement of $P_{g}(t) $ following this method, which matches the evolution obtained from the effective (circles) Hamiltonian Eq.~(\ref{HamilEff}).

\paragraph{Nonclassical states and Dirac equation.}

Nonclassical states of the EM field are a paramount resource for quantum information processing. However their generation using all-optical devices is challenging due to the lack of strong nonlinearities. Through the strong coupling between a qubit and a confined EM field, circuit/cavity-QED technology provides a way to overcome these limitations~\cite{Deleglise,Martinis,Nori}. Here we show that our setup can be used to generate highly nonclassical states of the field mode, without requiring the projective measurement of the qubit and/or the ability to control accurately the qubit-field interaction strength.

\begin{figure}
\includegraphics[width=0.22\textwidth]{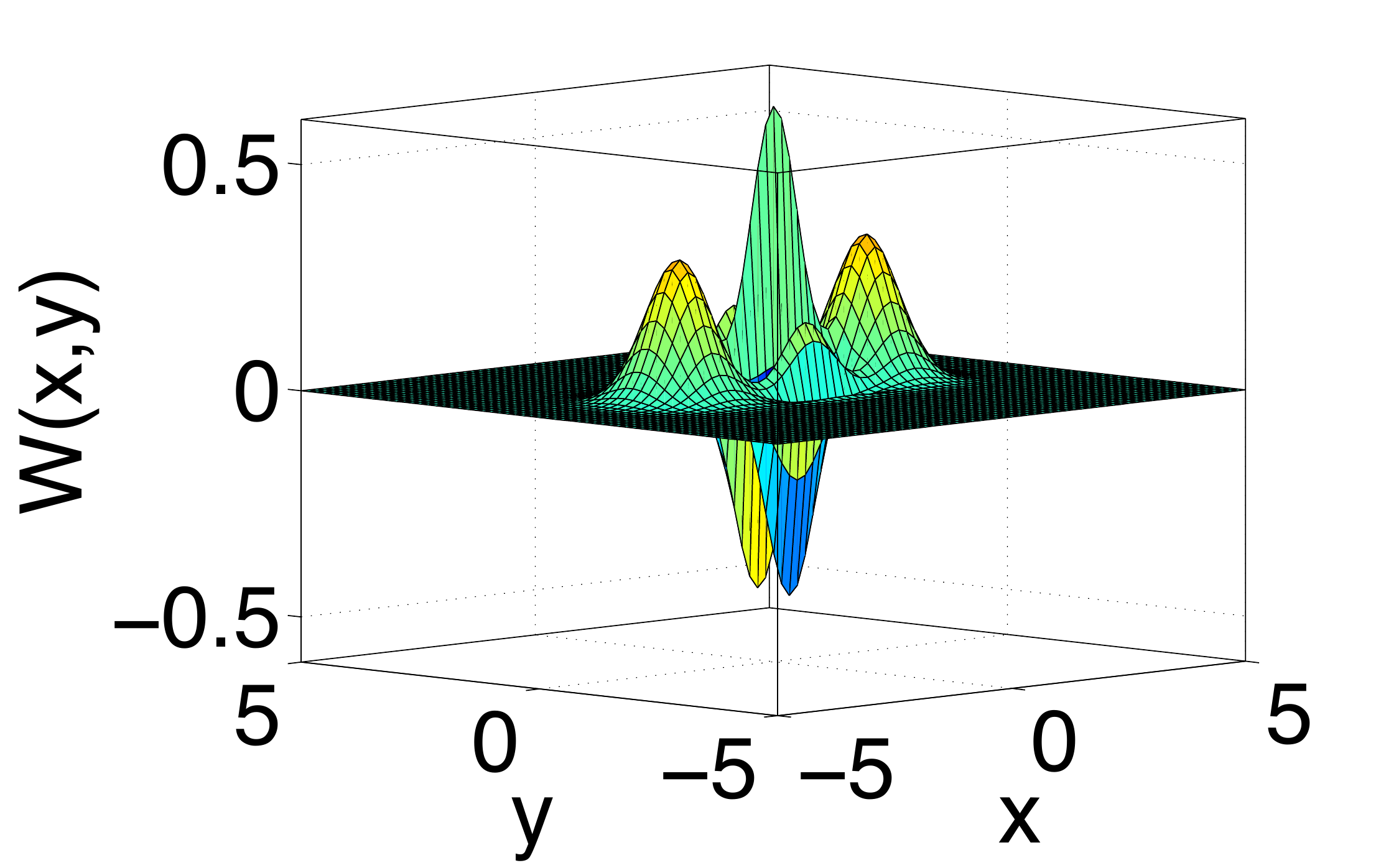}
\includegraphics[width=0.22\textwidth]{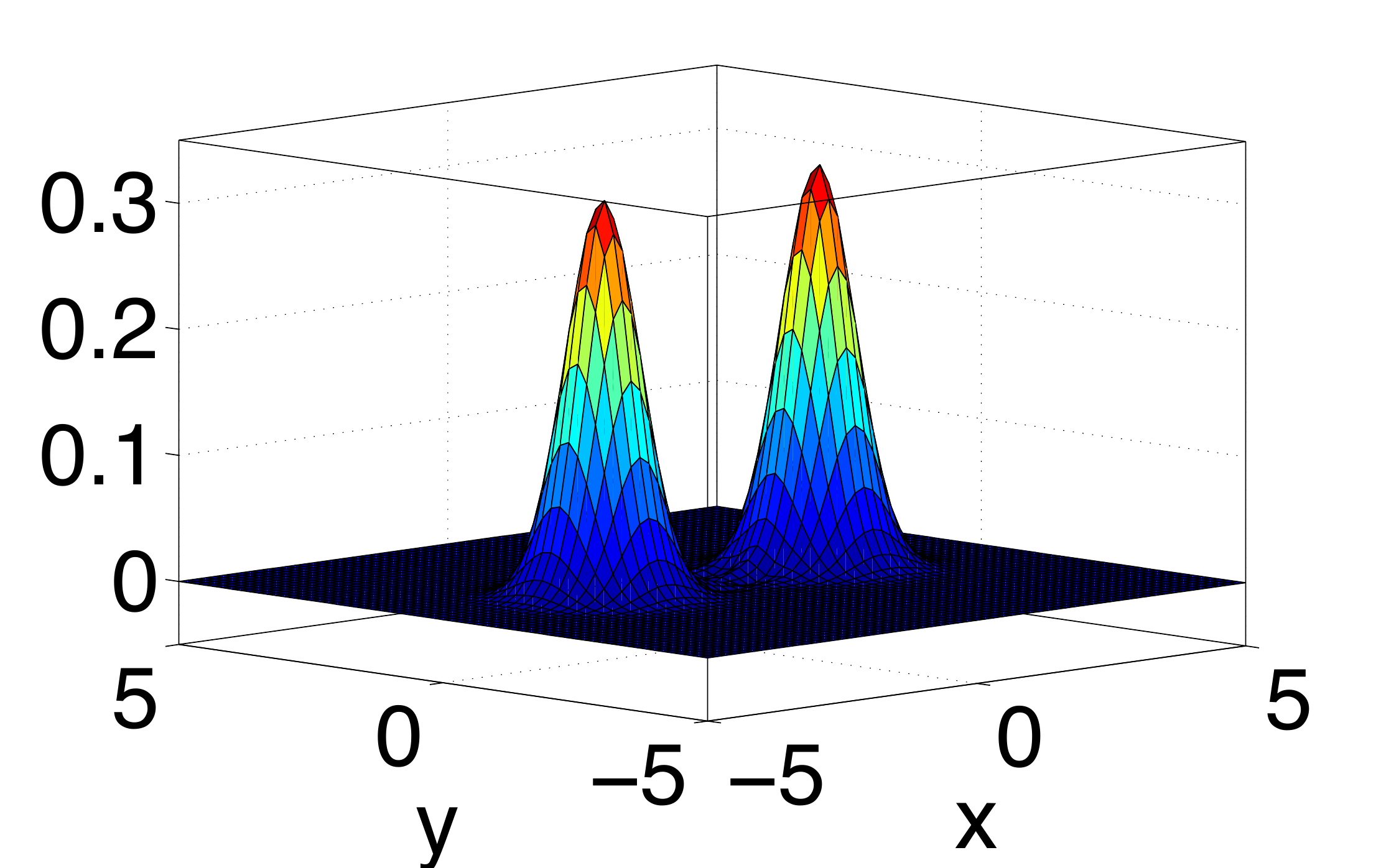}
\includegraphics[width=0.22\textwidth]{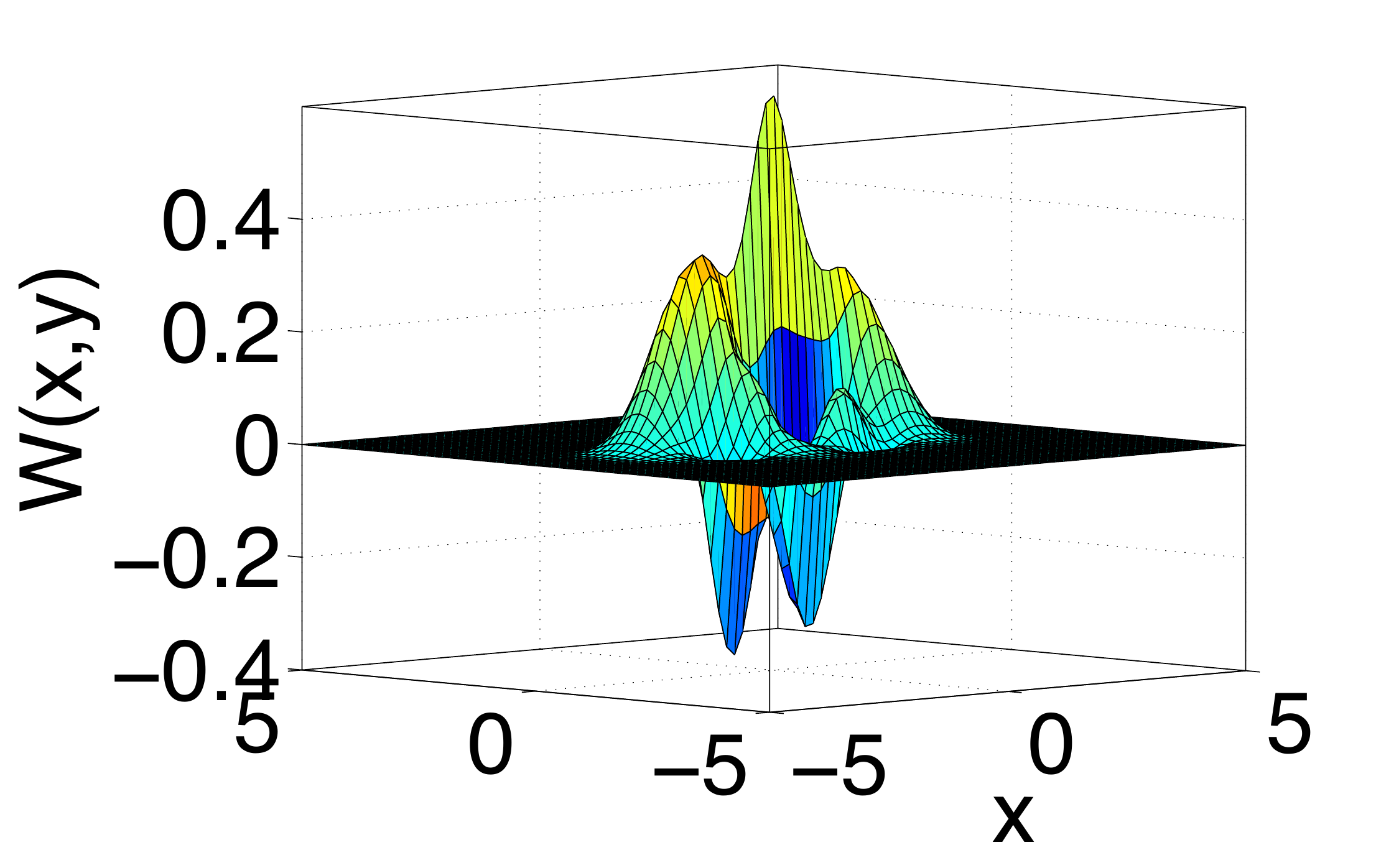}
\includegraphics[width=0.22\textwidth]{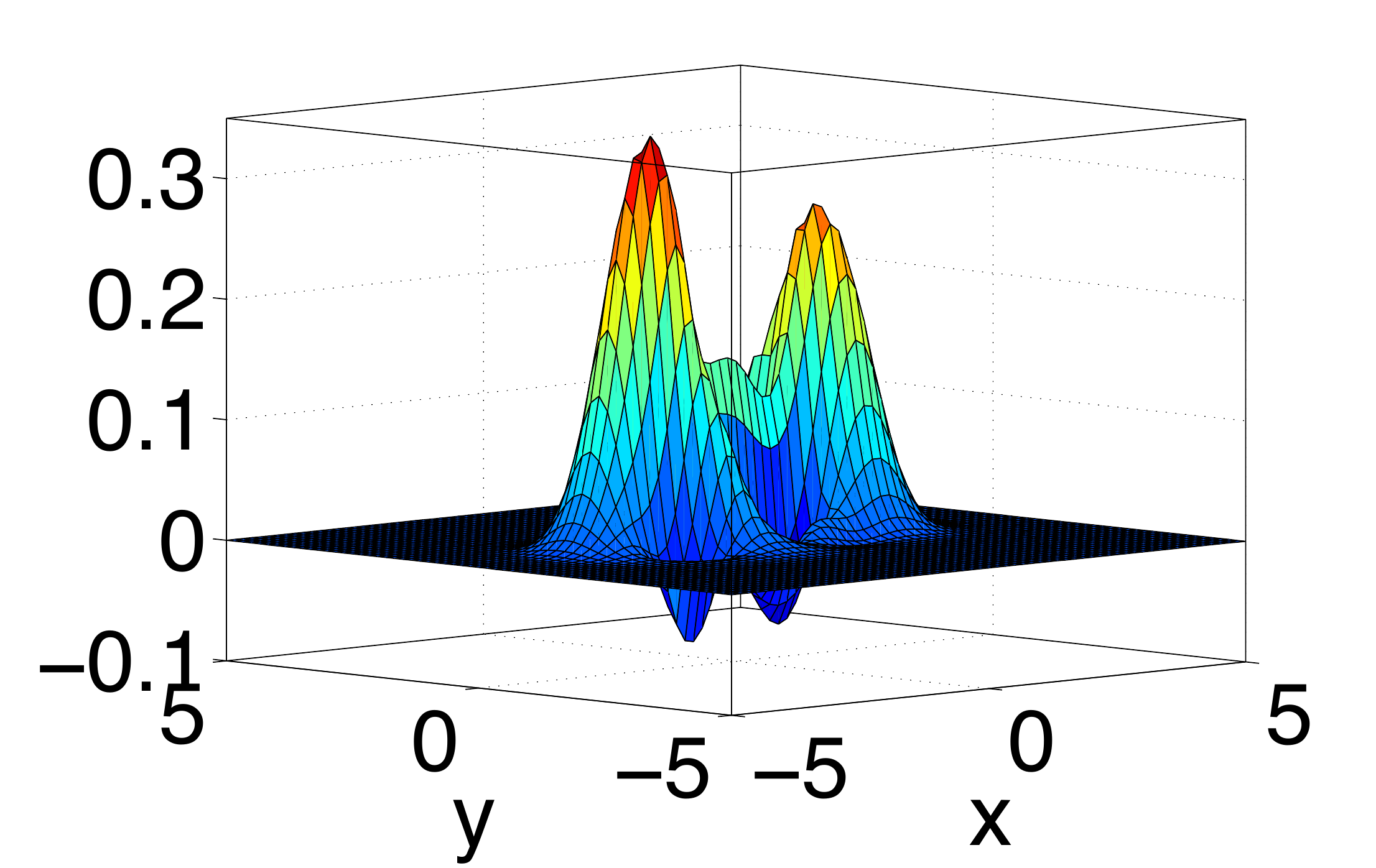}
\includegraphics[width=0.22\textwidth]{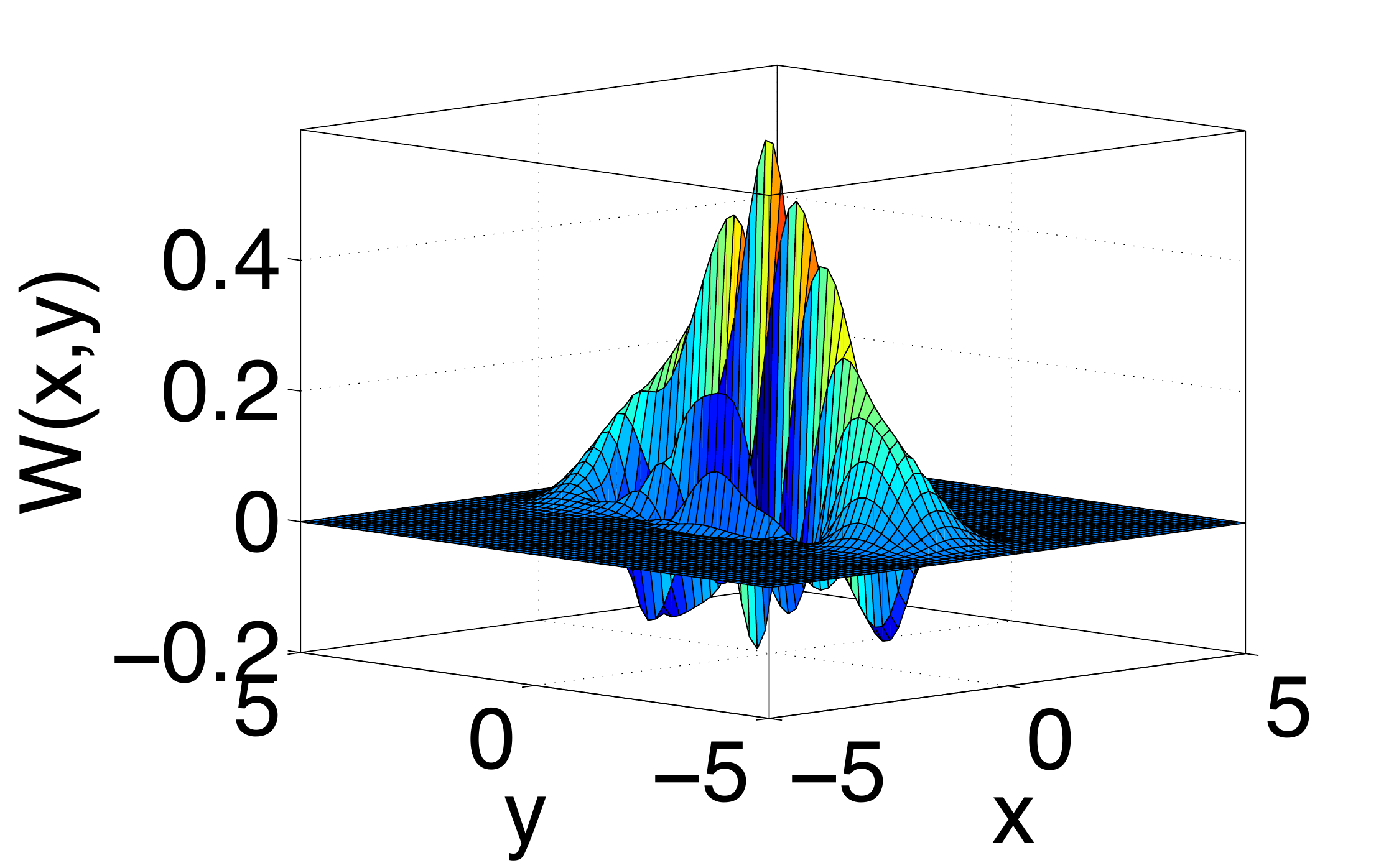}
\includegraphics[width=0.22\textwidth]{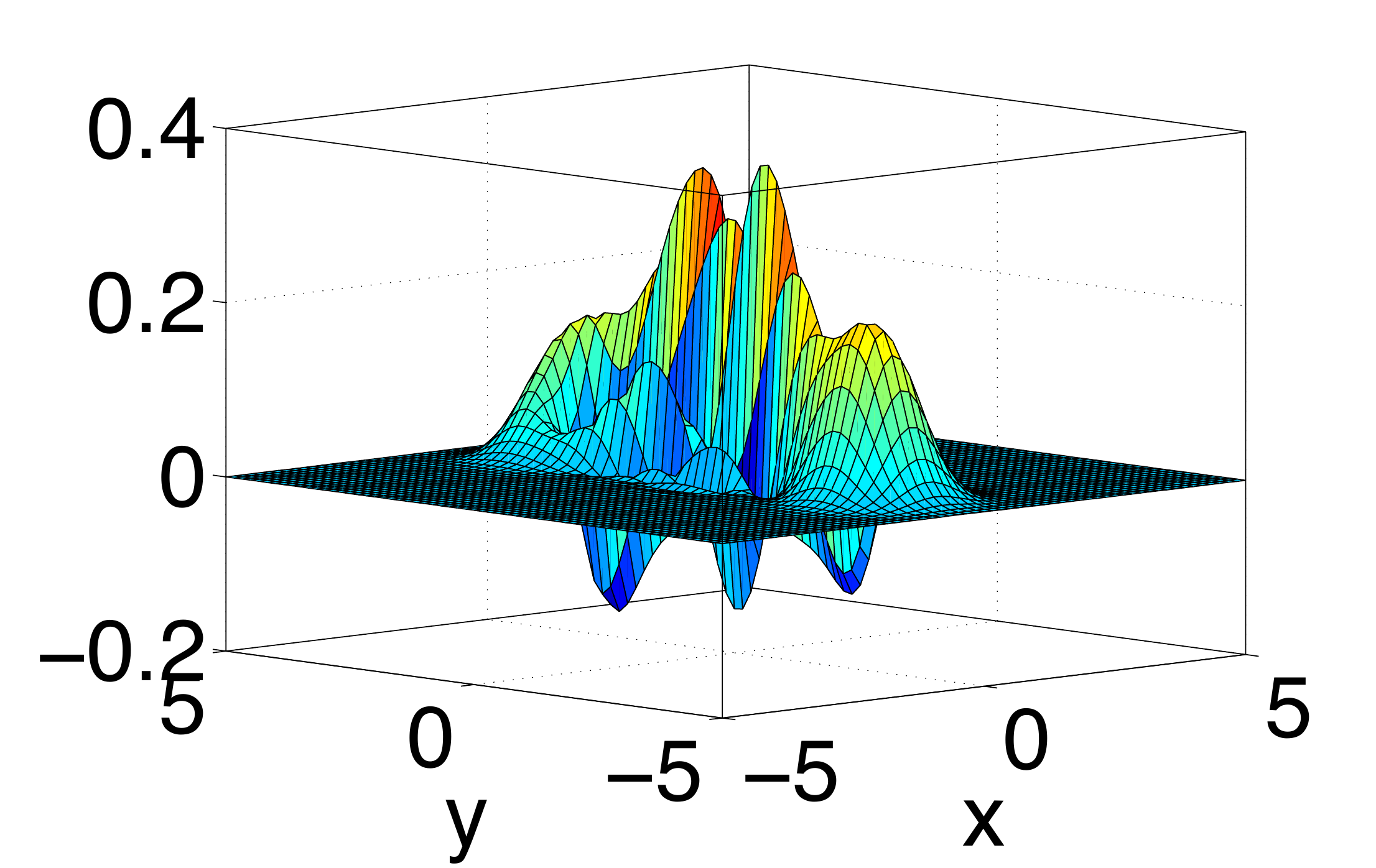}
\caption{\label{wigner:figs} Wigner function $W(x,y)$ of field state after interaction time $g_{\rm eff} t = \pi$, calculated ab initio with parameters of Fig.~\ref{cqed:figs}. {\bf (Top-left panel)} $\Omega_2 =0$, $\omega_{\rm eff}=g_{\rm eff}$, qubit postselected in ground state. {\bf (Top-right panel)} $\Omega_2  =0$, $\omega_{\rm eff}=g_{\rm eff}$, qubit traced out. {\bf (Center-left panel)} $\Omega_2  =2\pi\times10\,$MHz, $\omega_{\rm eff}=g_{\rm eff}$, qubit postselected in ground state. {\bf (Center-right panel)} $\Omega_2  =2\pi\times10\,$MHz, $\omega_{\rm eff}=g_{\rm eff}$, qubit traced out. {\bf (Bottom-left panel)} $\Omega_2  =2\pi\times10\,$MHz, $\omega_{\rm eff}=0$, qubit postselected in ground state. {\bf (Bottom-right panel)} $\Omega_2  =2\pi\times10\,$MHz, $\omega_{\rm eff}=0$, qubit traced out.} 
\end{figure}

The nonclassicality of a bosonic field can be signaled by the Wigner quasi-probability distribution function (WF), defined as
\begin{eqnarray}
W(\alpha)=\frac{2}{\pi} {\rm  Tr}\pare{D^\dag(\alpha) \rho_f D(\alpha) (-1)^{a^\dag a} } , \label{WF}
\end{eqnarray}
$\rho_{f}$ being the field density matrix and $D(\alpha)=\exp(\alpha a^\dag - \alpha^* a)$ the coherent displacement operator with amplitude $\alpha$. To show the ability of our setup to produce nonclassical states of the EM field, we plot in Fig.~\ref{wigner:figs} the WF of the field for the same set of parameters studied before, being the evolution time set at $g_{\rm eff} t = \pi$. The top-left panel on Fig.~\ref{wigner:figs} depicts the WF of the state generated when $\Omega_2  =0$ and the qubit is measured in its ground state. The result is a well known coherent Schr\"odinger-cat state with time-dependent amplitude $( g_{\rm eff}/\omega_{\rm eff} ) ( e^{-i \omega_{\rm eff}t} -1)$, which is nonclassical. However when the qubit is not measured, the state of the field after tracing out over the qubit will have as WF the plot of top-right panel in Fig.~\ref{wigner:figs}, where a classical mixture of coherent states with opposite phases is left and any quantum feature is erased. When $\Omega_2 \neq0$, a surprising property shows up after the qubit is traced out: the nonclassicality of the field is not completely lost (center-right panel on Fig.~\ref{wigner:figs}). This effect is more evident when $\omega_{\rm eff}=0$ (bottom-right panel on Fig.~\ref{wigner:figs}). The setup might be enhanced by taking advantage of the cavity output and novel measurement techniques~\cite{Menzel}, to produce nonclassical propagating microwaves and lasing in circuit QED~\cite{lasing}.

The case of $\omega_{\rm eff}=0$ is physically most relevant, as it relates to the quantum simulation of the 1+1 Dirac equation~\cite{Dirac}. If one repeats the derivation from Eq.~(\ref{HamilDriv}) but assumes that both external driving come with an additional phase $\phi=\pi/2$, i.e. $- \hbar \Omega_j  \pare{ e^{i (\omega_j t +\phi)} \sigma + {\rm H.c.}}$, and uses the rotated qubit basis $\ket{\pm_{\phi}} = \pare{\ket{g} \pm e^{-i \phi} \ket{e} }/\sqrt2$, then ${\cal H}_{\rm eff}$ becomes
\begin{eqnarray}
{\cal H}_{\rm D}
=   \frac{\hbar\Omega_2}{2} \sigma_z   +  \frac{\hbar g}{\sqrt2} \sigma_y \hat{p} , \label{HamilD}
\end{eqnarray}
for $\omega=\omega_1$, $\sigma_y=i(\sigma-\sigma^\dag)$, $\hat{p} =-i(a-a^\dag)/\sqrt2$. In the appropriate representation, the Schr\"odinger equation of our system is now formally equivalent to that of the 1+1 Dirac equation of a relativistic spin-1/2 particle, 
\begin{eqnarray}
i\hbar \frac{d \psi}{dt}
=  \pare{ c \hat{p}  \sigma_y + mc^2 \sigma_z   } \psi , \label{Dirac}
\end{eqnarray}
where $\hat{p}$ represents the momentum of a Dirac particle of mass $m$ and $c$ the speed of light. This shows the ability of our proposal to access a wide range of physical models. Through the engineered Hamiltonian, a qubit at rest coupled to the EM field would behave as a massive spin-1/2 particle moving near the speed of light. To observe peculiar effects, such as {\it Zitterbewegung}, in our setup one must pay attention at the mapping of the bosonic degree of freedom. While in the original Dirac equation (\ref{Dirac}) the operator $\hat{p}$ corresponds to the momentum of the Dirac particle, in our Hamiltonian Eq. (\ref{HamilD}) this role is played by the quadrature $\hat{p} =-i(a-a^\dag)/\sqrt2$ of the EM field mode in the resonator. Hence, in the simulated Dirac equation the dynamics of {\it Zitterbewegung} will show up in the expectation value of the field quadrature $\hat{x} =(a+a^\dag)/\sqrt2$, which has the same time evolution as the expectation value of the position operator $\hat{x}$ of a Dirac particle.

\paragraph{Discussion and conclusions.}

Although we have disregarded the possible coupling between the orthogonal driving and the resonator field, it is easy to show that the effect of such a spurious coupling could be sorted out by adding a driving to the cavity (acting as a counter coherent displacement). Another source of error stems from qubit dephasing and relaxation rates, as well as the cavity decay rate. Nonetheless, interaction times considered in all numerical simulations are well below standard decoherence times. To avoid excitation of higher levels in the qubit, it is possible to design flux qubits where the splitting to the third level is above 30 GHz~\cite{Niemczyk, FD}. Our method can be extended to the case of a slightly anharmonic qubit via Gaussian shaped DRAG pulses~\cite{pulses}.

The proposed quantum simulation of a broad range of regimes of light-matter coupling may become a building block to simulate physics inaccessible in standard quantum optics. This includes the Dicke model for multiple qubits, the spin-boson model when the resonator is replaced by an open transmission line, Jahn-Teller instability for several discrete modes in the resonator, and relativistic quantum physics such as scattering of relativistic particles \cite{Lucas}. Efficient computations of these problems may be beyond current numerical capabilities.

We thank Jean-Michel Raimond for insightful discussions about this work. The authors acknowledge funding from Juan de la Cierva MICINN Program, Spanish projects MICINN FIS2009-10061 and FIS2009-12773-C02-01, UPV/EHU UFI 11/55, QUITEMAD, Basque Government IT472-10, SOLID, CCQED, and PROMISCE European projects, the SFB 631 of the Deutsche Forschungsgemeinschaft and the German Excellence Initiative via NIM.

\end{document}